\documentclass[preprint,showpacs,preprintnumbers,amsmath,amssymb]{revtex4}

\usepackage{graphicx}
\usepackage{dcolumn}
\usepackage{bm}
\usepackage{hyperref}
\usepackage{array}
\hypersetup{colorlinks=False}

\begin{document}
\newcommand{\RR}[1]{[#1]}
\newcommand{\intsum}{\sum \kern -15pt \int}
\newfont{\Yfont}{cmti10 scaled 2074}
\newcommand{\Y}{\hbox{{\Yfont y}\phantom.}}
\def\O{{\cal O}}
\newcommand{\bra}[1]{\left< #1 \right| }
\newcommand{\braa}[1]{\left. \left< #1 \right| \right| }
\def\Bra#1#2{{\mbox{\vphantom{$\left< #2 \right|$}}}_{#1}
\kern -2.5pt \left< #2 \right| }
\def\Braa#1#2{{\mbox{\vphantom{$\left< #2 \right|$}}}_{#1}
\kern -2.5pt \left. \left< #2 \right| \right| }
\newcommand{\ket}[1]{\left| #1 \right> }
\newcommand{\kett}[1]{\left| \left| #1 \right> \right.}
\newcommand{\scal}[2]{\left< #1 \left| \mbox{\vphantom{$\left< #1 #2 \right|$}}
\right. #2 \right> }
\def\Scal#1#2#3{{\mbox{\vphantom{$\left<#2#3\right|$}}}_{#1}
{\left< #2 \left| \mbox{\vphantom{$\left<#2#3\right|$}} \right. #3
\right> }}


\title{New insight into $nd\rightarrow$ $^3H\gamma$ process at thermal energy with pionless effective field theory}

\author{M. Moeini Arani}%
\email[]{m.moeini.a@khayam.ut.ac.ir (corresponding author)}
\affiliation {Department of Physics, University of Tehran, P.O.Box
14395-547, Tehran, Iran }
\author{H. Nematollahi}
\email[]{hnematollahi.90@ut.ac.ir}%
\affiliation {Department of Physics, University of Tehran, P.O.Box
14395-547, Tehran, Iran }
\author{N. Mahboubi }
\email[]{n.mahboubi@ut.ac.ir} \affiliation {Department of Physics,
University of Tehran, P.O.Box 14395-547, Tehran, Iran }
\author{S. Bayegan }
\email[]{bayegan@khayam.ut.ac.ir} \affiliation {Department of
Physics, University of Tehran, P.O.Box 14395-547, Tehran, Iran }%

\date{\today}

\begin{abstract}
We take a new look at the neutron radiative capture by a deuteron at
thermal energy with the pionless effective field theory
(EFT($\pi\!\!\!/$)) approach. We present in detail the calculation
of $nd\rightarrow$ $^3H\gamma$ amplitudes for incoming doublet and
quartet channels leading to the formation of a triton fully in the
projection method based on the cluster-configuration space approach.
In the present work, we consider all possible one-body and two-body
photon interaction diagrams. In fact, additional diagrams that make
significant changes in the results of the calculation of the total
cross section in the $nd\rightarrow$ $^3H\gamma$ process are
included in this study. The properly normalized triton wave function
is calculated and taken into consideration. We compare the cross
section of the dominant magnetic M1-transition of $nd\rightarrow$
$^3H\gamma$ up to next-to-next-to-leading order
$\textrm{N}^2\textrm{LO}$ with the results of the previous
model-dependent theoretical calculations and experimental data. The
more acceptable results for cross section
$\sigma^{(2)}_{tot}=0.297\;(\textrm{LO})+0.124\;(\textrm{NLO})+0.048\;(\textrm{N}^2\textrm{LO})=[0.469\pm0.033]\:\textrm{mb}$
show order by order convergence and cutoff independence. No
three-body currents are needed to renormalize observables up to
$\textrm{N}^2\textrm{LO}$ in this process.
\end{abstract}

\maketitle

\section{Introduction}\label{sec:introduction}
Studies of the radiative capture reactions on numerous light atomic
nuclei have been continued at thermal and astrophysical energies
with the model-independent pionless effective field theory
(EFT($\pi\!\!\!/$)) approach in the recent years
\cite{Rupak-np}-\cite{ Acharya-phillips}.

The calculation  of radiative capture amplitude and cross section of
$nd\rightarrow$ $^3H\gamma$ and $pd\rightarrow$ $^3H\!e\,\gamma$ are
an essential input in the calculation of the parity-violating
radiative capture of the above processes at the thermal energy
\cite{Moskalev,desplanques-benayoun,moeini-bayegan}.

In the present paper, we study the $nd\rightarrow$ $^3H\gamma$
process fully with the projection operator method based on the
cluster-configuration space which is introduced by \cite{20 of
sadeghi-bayegan}. We also consider the calculation of observables
with M1 transition up to next-to-next-to-leading order
($\textrm{N}^2\textrm{LO}$) with the following significant changes
in comparison with the previous EFT($\pi\!\!\!/$) calculation
\cite{sadeghi-bayegan-grieshammer}: a) including the diagrams with
radiation from external nucleon leg, external deuteron leg, and
on-shell two-body bubble (see the diagrams "$a_0$", "$a_1$", and
"$a_3$" in Fig.\ref{Fig: nd capture} of page \pageref{Fig: nd
capture}), b) considering both contributions corresponding to two
nucleon poles before and after photon creation in the first diagram
of the second row in Fig.\ref{Fig: nd capture}, c) inserting the
diagram with the radiation directly from the exchanged nucleon, d)
adding the contribution of the $^3S_1\rightarrow^3S_1$ M1 transition
and e) introducing .and using the properly normalized triton wave
function.

The triton or helium-three wave functions consist of two parts, one
is the nucleon and dibaryon cluster wave function and the other is
the two nucleon structure of the dibaryon cluster. We follow the
Bethe-Salpeter (BS) equation in \cite{adam,book for normalization}
and we use the normalization condition of the relativistic two-body
vertex function and work out the nonrelativistic one which is
suitable for neutron-deuteron ($nd$) scattering leading to the
formation of a triton.

The theoretical calculations of the observables in the
$nd\rightarrow$ $^3H\gamma$ process were previously performed based
on model-dependent approaches \cite{Faldt et al}. The cross section
and polarization observables were studied theoretically for
radiative capture reactions $^2H$$(n,\gamma)$$^3H$ and
$^2H$$(p,\gamma)$$^3He$ at low energies \cite{viviani et al}. The
cross section for thermal neutron radiative capture on the deuteron
was measured to be $\sigma_{tot}^{exp}=0.508\pm0.015$ mb
\cite{Jurney et al}, in agreement with the results of earlier
experiments \cite{Kaplan et al,Merritt et al}.

In the present work, the calculation of all M1 diagrams are
calculated for the incoming doublet and quartet $nd$ channels fully
in the cluster-configuration space up to $\textrm{N}^2\textrm{LO}$
in Sec.\ref{sec:amplitude}. The calculation of the cross section for
$nd\rightarrow$ $^3H\gamma$ is presented in Sec.\ref{Sec:cross
section}. In Sec.\ref{sec:numerical} numerical aspects of the
calculation of M1 amplitudes are discussed. The results and
comparison with other theoretical and experimental works are
explained in Sec.\ref{sec:results}. Finally, we summarize the paper
and discuss future investigations in Sec.\ref{sec:conclusion}.

\section{$nd\rightarrow$ $^3H\gamma$ system}\label{sec:amplitude}
In this section, we focus on the introduction of the
EFT($\pi\!\!\!/$) amplitude for the $nd\rightarrow$ $^3H\gamma$
process up to $\textrm{N}^2\textrm{LO}$. We concentrate on the
zero-energy regime and try to calculate the amplitude of the neutron
radiative capture by deuteron at thermal energy ($2.5\times10^{-8}$
MeV) in the center-of-mass (c.m.) frame.

In the EFT($\pi\!\!\!/$) method, the electromagnetic (EM)
interactions in the three-body systems can be inserted principally
by considering the one-, two-, and three-body currents. However, we
show that the cutoff independence is achieved up to
$\textrm{N}^2\textrm{LO}$ with one- and two-body currents and
therefore there is no need for additional three-body currents up to
$\textrm{N}^2\textrm{LO}$ calculations. In the very-low-energy
regime the M1 transition has a dominant piece in the amplitude of
$nd\rightarrow$ $^3H\gamma$. E2 transition also contribute to the
$nd\rightarrow$ $^3H\gamma$ reaction but comparing with the M1
interaction, it has a negligible contribution. In the following, we
evaluate the EFT($\pi\!\!\!/$) amplitude of the neutron radiative
capture by deuteron reaction by considering the dominant M1
transitions using one- and two-body currents up to
$\textrm{N}^2\textrm{LO}$.

Note that the convection current of the proton (E1 transition) has
odd parity (due to one power of nucleon momentum), so this mixes an
incoming P-wave state to the final S-wave triton. Capture from the P
wave introduces the factor of the external nucleon momentum forcing
the amplitude to vanish at threshold.

The Lagrangian of the S-wave strong interactions using a dibaryon
auxiliary field are given by \cite{phillips-rupak-savage,20 of
sadeghi-bayegan}
\begin{eqnarray}\label{Eq:1}
\mathcal{L}_{S}=N^\dag\Big(iD_0+\frac{\vec{D}^2}{2m_N}\Big)N+d^{A^\dag}_{s}\Big[\Delta_s-c_{0s}\Big(iD_0+\frac{\vec{D}^2}{4m_N}+\frac{\gamma^2_s}{m_N}\Big)\Big]d^{A}_{s}
  \qquad\qquad\qquad\qquad\quad\nonumber \\
  +d^{i^\dag}_{t}\Big[\Delta_t-c_{0t}\Big(iD_0+\frac{\vec{D}^2}{4m_N}+\frac{\gamma^2_t}{m_N}\Big)\Big]d^{i}_{t}
  -y\Big(d^{A^\dag}_{s}(N^\dag P^A N)+d^{i^\dag}_{t}(N^\dag P^i
  N)+h.c.\Big)\quad\!\!\!\nonumber\\ +\frac{m_N y^2
  \mathcal{H}(E,\Lambda)}{6}\;N^\dag\bigg(\big(d^{i}_{t}\sigma_i\big)^\dag\big(d^{j}_{t}\sigma_j\big)-\big[\big(d^{i}_{t}\sigma_i\big)^\dag\big(d^{A}_{s}\sigma_A\big)
  +h.c.\big]+\big(d^{A}_{s}\tau_A\big)^\dag\big(d^{B}_{s}\sigma_B\big)\bigg)N\!\!\!\!\!\!\!\!\nonumber\\+\cdot\cdot\cdot\,,\;\quad\qquad\qquad\qquad\qquad\qquad\qquad\qquad\qquad\qquad\qquad\qquad\qquad\qquad\qquad\qquad\:\;
\end{eqnarray}
where $D_\mu$ is the covariant derivative which acts on the nucleon
and dibaryon fields with $
\partial_\mu + ie\frac{1+\tau_3}{2}A_\mu$ and
$\partial_\mu + ieCA_\mu$ relations, respectively. $A_\mu$ is the
external field and $C=2,1,$ and $0$ for proton-proton,
neutron-neutron and neutron-neutron dibaryons. The center dots in
the last line denotes the other suppressed terms. In
Eq.(\ref{Eq:1}), $N$ is the nucleon iso-doublet field. The dibaryon
auxiliary fields for deuteron and iso di-nucleon systems are
introduced by $d^{i}_{t}$ and $d^{A}_{s}$, respectively. The
operators $P^{i}=\frac{1}{\sqrt{8}}\sigma_2\sigma^i\tau_2$ and
$P^{A}=\frac{1}{\sqrt{8}}\sigma_2\tau_2\tau^A$ with $\tau_{A}$
($\sigma_{i}$) as isospin (spin) Pauli matrices are the projection
operators of nucleon-nucleon ($NN$) $^3S_1$ and $^1S_0$ states,
respectively. $m_N$ represents the nucleon mass and the
three-nucleon force is introduced by $\mathcal{H}(E,\Lambda)$, where
$E$ and $\Lambda$ are the total energy and cutoff momentum. The
$\mathcal{H}(E,\Lambda)$, which absorbs all dependence on the cutoff
as $\Lambda\rightarrow\infty$, is given by
\cite{Bedaque-H-vK,Bedaque-H-vK-2,Bedaque-R-H}
\begin{eqnarray}\label{Eq:05}
  \mathcal{H}(E,\Lambda)=\frac{2}{\Lambda^2}\sum_{m=0}^{\infty}H_{2m}(\Lambda)\bigg(\frac{m_N
  E+\gamma_t^2}{\Lambda^2}\bigg)^m=\frac{2H_0(\Lambda)}{\Lambda^2}+\frac{2H_2(\Lambda)}{\Lambda^4}(m_N
  E+\gamma_t^2)+\cdot\cdot\cdot\,,
\end{eqnarray}
where the interactions proportional to $H_{2m}$ enter at
$\textrm{N}^{2m}\textrm{LO}$ \cite{Bedaque-R-H}.

In our calculation, we consider generally $y^2=\frac{4\pi}{m_N}$.
The parameters $\Delta_{s/t}$ and $c_{0s/t}$ are given by matching
the EFT($\pi\!\!\!/$) $NN$ scattering amplitude to the effective
range expansion (ERE) of the scattering amplitude of two
non-relativistic nucleons around the $i\gamma_{s/t}$ \cite{20 of
sadeghi-bayegan}. $\gamma_t=45.7025$ MeV is the binding momentum of
the deuteron and $\gamma_s=\frac{1}{a_s}$ with $a_s=-23.714$ fm as
the scattering length in the $^1S_0$ state.

The Lagrangian of the M1 interaction is constructed by considering
the nucleon and dibaryon operators coupling to the magnetic field
$\vec{B}$,
\begin{eqnarray}\label{Eq:6}
  \mathcal{L}_B=\frac{e}{2m_N}N^\dag\big(k_0+k_1\tau^3\big)\vec{\sigma}\cdot\vec{B}N+e\frac{L_1}{m_N\sqrt{\rho_d
  r_0}}d^{j^\dag}_{t}d^{3}_{s}B_j-e\frac{2L_2}{m_N\rho_d}i\,\epsilon_{ijk}\,d^{i^\dag}_{t}d^{j}_{t}B_k+h.c.\,.
\end{eqnarray}
In the above equation, $k_0=\frac{1}{2}(k_p+k_n)=0.4399$ and
$k_1=\frac{1}{2}(k_p-k_n)=2.35294$ with $k_p$ ($k_n$) as the proton
(neutron) magnetic moment are the isoscalar and isovector nucleon
magnetic moments, respectively. $e$ is the electric charge and
$\rho_d=1.764$ fm ($r_0=2.73$ fm) denotes the effective range of the
triplet (singlet) $NN$ state. The coefficients $L_1=-4.427\pm0.015$
fm and $L_2=-0.4$ fm, which enter at next-to-leading order (NLO),
have been fixed from the cross section of $np\rightarrow$ $d\gamma$
at thermal energy,
$\sigma^{exp}_{np\rightarrow\,d\gamma}=334.2\pm0.5 $ mb and the
deuteron magnetic moment $\mu_M$, respectively \cite{Ando-Hyun}.

\begin{figure}
\includegraphics*[width=16.5cm]{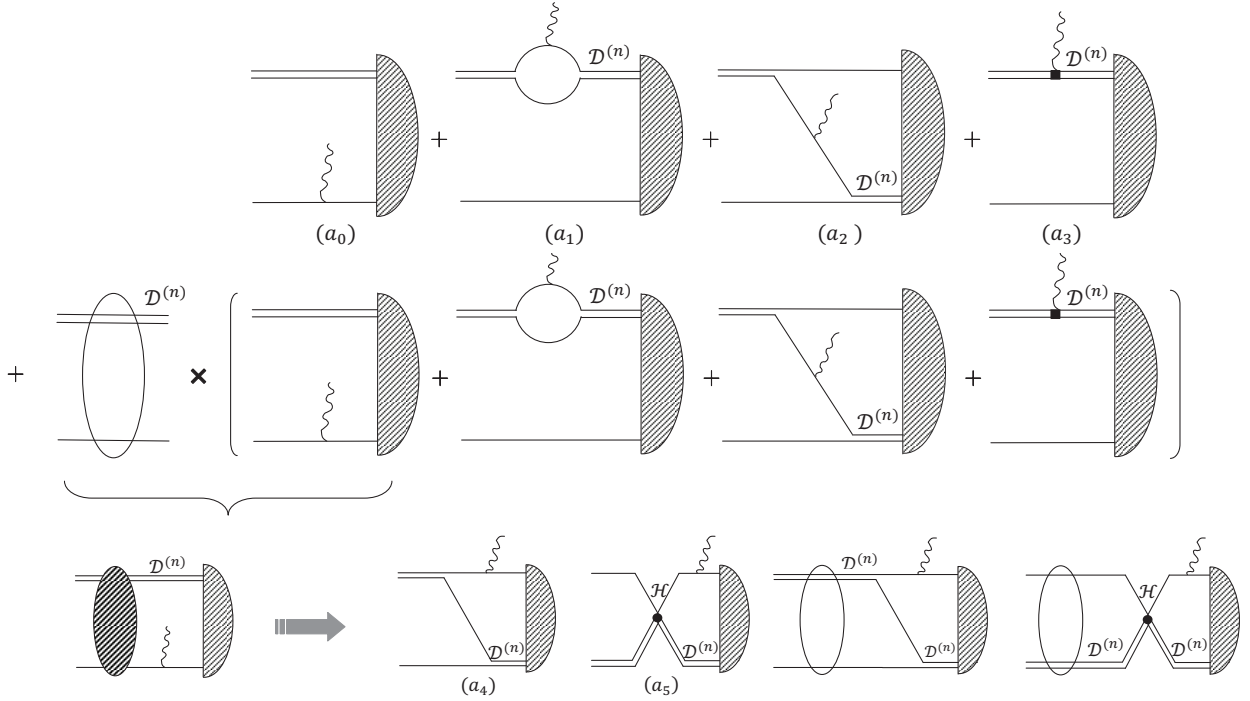}\centering
\caption{\label{Fig: nd capture}The M1 $nd\rightarrow$ $^3H\gamma$
diagrams at $\textrm{N}^n\textrm{LO}$ ($n\leq2$). The "$^{(n)}$"
superscript denotes the contribution up to
$\textrm{N}^n\textrm{LO}$. All possible diagrams in the M1
transition of the $nd\rightarrow$ $^3H\gamma$ process up to
$\textrm{N}^n\textrm{LO}$ are shown in the first and second lines.
The diagrams in the third line are the expanded version of the first
diagram in the second line. The solid, wavy, and double lines
represent a nucleon, a photon, and a dibaryon field, respectively.
The nucleon-nucleon-photon ($NN\gamma$) and dibaryon-dibaryon-photon
($dd\gamma$) vertices show the one- and two-body M1 interactions.
$\mathcal{D}^{(n)}$ is the $2\times 2$ propagator matrix of the
dibaryon auxiliary fields and the three-body force is indicated by
$\mathcal{H}$. The dashed oval and dashed half oval denote the $Nd$
scattering amplitude and the normalized triton wave function up to
$\textrm{N}^n\textrm{LO}$, respectively.}
\end{figure}
The diagrams of the M1 transition in the $nd\rightarrow$ $^3H\gamma$
process up to $\textrm{N}^n\textrm{LO}$ ($n\leq2$) are schematically
shown in Fig.\ref{Fig: nd capture}. Note that in the entire paper,
the superscript "(n)" denotes the contribution from the sum of all
pieces up to, and including, order n. $\mathcal{D}^{(n)}$ indicates
the propagator of the dibaryon fields up to
$\textrm{N}^n\textrm{LO}$ which is given in the
cluster-configuration space as
\begin{eqnarray}\label{Eq:06}
\mathcal{D}^{(n)}(q_0,q)=\left(
                           \begin{array}{cc}
                             D_t^{(n)}(q_0-\frac{q^2}{2m_N},q) & 0 \\
                             0 & D_s^{(n)}(q_0-\frac{q^2}{2m_N},q) \\
                           \end{array}
                         \right),
\end{eqnarray}
where
\begin{eqnarray}\label{Eq:006}
D^{(n)}_t(q_0,q)=\frac{1}{\gamma_{t}-\sqrt{\frac{q^2}{4}-m_Nq_0-i\varepsilon}}\sum^{n}_{m=0}\bigg(\frac{\frac{\rho_d}{2}(m_N q_0-\frac{q^2}{4}+\gamma_t^2)}{\gamma_{t}-\sqrt{\frac{q^2}{4}-m_Nq_0-i\varepsilon}}\bigg)^m,\nonumber \\
D^{(n)}_s(q_0,q)=\frac{1}{\gamma_{s}-\sqrt{\frac{q^2}{4}-m_Nq_0-i\varepsilon}}\sum^{n}_{m=0}\bigg(\frac{\frac{r_0}{2}(m_N
q_0-\frac{q^2}{4})}{\gamma_{s}-\sqrt{\frac{q^2}{4}-m_Nq_0-i\varepsilon}}\bigg)^m.
\end{eqnarray}
We emphasize that the above propagators can be applied up to
$\textrm{N}^3\textrm{LO}$ and should be corrected for the higher
orders.

In Fig.\ref{Fig: nd capture}, the dashed oval denotes the
nucleon-deuteron ($Nd$) scattering amplitudes which are presented by
$t^{(n)}_{d}$ and $t^{(n)}_{q}$ for the doublet and quartet channels
up to $\textrm{N}^n\textrm{LO}$, respectively. The Faddeev equations
of $t^{(n)}_{d/q}$ are introduced in Appendix \ref{Appendix A}. The
dashed half oval indicates the normalized triton wave function up to
$\textrm{N}^n\textrm{LO}$ which is introduced by $t^{(n)}_{^3H}$ in
the following. The procedure of making the triton wave function and
its normalization condition are briefly presented in Appendix
\ref{Appendix B}.

We consider the contribution of all diagrams shown in Fig.\ref{Fig:
nd capture} in the amplitude of neutron radiative capture by a
deuteron reaction. The third diagram of the second line and all
diagrams of the first line in Fig.\ref{Fig: nd capture} have not
been considered in the previous EFT($\pi\!\!\!/$) calculations of
the $nd\rightarrow$ $^3H\gamma$ amplitude
\cite{sadeghi-bayegan,sadeghi-bayegan-grieshammer}. We have also
added the contribution of the $^3S_1\rightarrow^3$$S_1$ M1
transition to the amplitude of the $nd\rightarrow$ $^3H\gamma$
process which was previously not considered in
\cite{sadeghi-bayegan,sadeghi-bayegan-grieshammer}. This two-body M1
transition is indicated in the Lagrangian of Eq.(\ref{Eq:6}) by the
$L_2$ coefficient which enters first at NLO as $L_1$. However the
contribution of the $^3S_1\rightarrow^3$$S_1$ M1 transition is small
at NLO but its effect is significant at $\textrm{N}^2\textrm{LO}$.

Before we evaluate the contribution of the diagrams in Fig.\ref{Fig:
nd capture}, let us make a comment about the computational process
of the amplitude at NLO and $\textrm{N}^2\textrm{LO}$. Introducing
the $\textrm{N}^n\textrm{LO}$ diagrams as in Fig.\ref{Fig: nd
capture}, includes some diagrams of higher order, for example the
NLO calculation includes $\textrm{N}^2\textrm{LO}$,
$\textrm{N}^3\textrm{LO}$ and $\textrm{N}^4\textrm{LO}$ terms. So,
this calculation includes higher-order terms, but it is not - or at
least, not immediately - a full higher-order correction, and so does
not achieve that precision. On the other hand, the additional
diagrams are small in a well behaved expansion, so the precision is
not compromised. This procedure is made only for convenience in the
computational process.

Now, we make a comment about the evaluation of the first diagram in
the second line of Fig.\ref{Fig: nd capture}. This diagram is
different somewhat from other ones because it has two contributions
corresponding to the poles in the nucleon propagators before and
after the photon creation. Therefore, these two poles are
corresponding to two contributions, one is that the photon is
emitted during the exchange of a nucleon and the other is that the
photon is emitted after exchanging the nucleon. If we add the
half-offshell $Nd$ scattering amplitude from left to the first
diagram in the first line of Fig.\ref{Fig: nd capture}, we miss the
contribution of the case that the photon emission occurs during the
nucleon exchange. So, we have to replace the half-offshell $Nd$
scattering amplitude by four diagrams which are introduced in
Fig.\ref{Fig:nd scattering}. Thus, we must substitute the first
diagrams of the second line by four diagrams introduced in the third
line of Fig.\ref{Fig: nd capture}. The effect of the photon emitted
during the exchange of another nucleon can also be applied when the
triton formation precedes the photon-nucleon interaction. But the
evaluation of this effect makes no significant changes in the final
results.

By working in the Coulomb gauge, the M1 amplitude of the
$nd\rightarrow$ $^3H\gamma$ can be written as two orthogonal terms,
\begin{eqnarray}\label{Eq:7}
  \big(t^\dag\sigma_aN\big)\big(\vec{\varepsilon}_d\times\big(\vec{\varepsilon}^\ast_\gamma\times\vec{\tilde{q}}\big)\big)_a\:,\qquad i\big(t^\dag
  N\big)\big(\vec{\varepsilon}_d\cdot\vec{\varepsilon}^\ast_\gamma\times\vec{\tilde{q}}\big)\,,\;\;\;
\end{eqnarray}
with $t$, $\vec{\varepsilon}_\gamma$, $\vec{\varepsilon}_d$, and
$\vec{\tilde{q}}$ are the final $^3H$ (or $^3He$) field, the
three-vector polarization of the produced photon, the three-vector
polarization of the deuteron and the unit vector along the
3-momentum of the photons, respectively.

In the $nd\rightarrow$ $^3H\gamma$ process, two initial doublet
($^2S_{\frac{1}{2}}$) and quartet ($^4S_{\frac{3}{2}}$) channels can
make the final triton state using the M1 transition. If we evaluate
the contributions of all diagrams in Fig.\ref{Fig: nd capture}, we
can generally write the $\textrm{N}^n\textrm{LO}$ ($n\leq2$)
amplitude of the $nd\rightarrow$ $^3H\gamma$ process as
\begin{eqnarray}\label{Eq:8}
  \mathcal{W}^{(n)}=t^\dag\big[\mathcal{M}^{(n)}_d Y_d+\mathcal{M}^{(n)}_q Y_q]N
\end{eqnarray}
where
\begin{eqnarray}\label{Eq:9}
  Y_d=i\vec{\varepsilon}_d\cdot\vec{\varepsilon}^\ast_\gamma\times\vec{\tilde{q}}+\vec{\sigma}\times\vec{\varepsilon}_d\cdot\vec{\varepsilon}^\ast_\gamma\times\vec{\tilde{q}}\,,\:\,
\nonumber\\
Y_q=2\,i\vec{\varepsilon}_d\cdot\vec{\varepsilon}^\ast_\gamma\times\vec{\tilde{q}}-\vec{\sigma}\times\vec{\varepsilon}_d\cdot\vec{\varepsilon}^\ast_\gamma\times\vec{\tilde{q}}\,.
\end{eqnarray}

\begin{figure}
\includegraphics*[width=6cm]{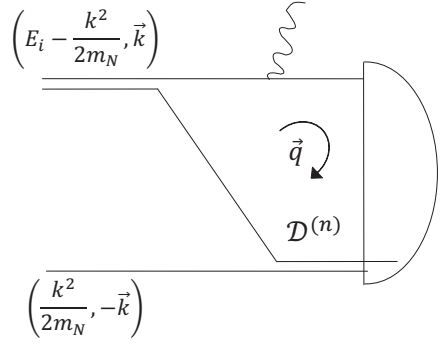}\centering
\caption{\label{Fig:a4}The $a_4$ diagram in Fig.\ref{Fig: nd
capture}. $k$ and $E_i$ denote the incoming c.m. momentum and the
total energy of the initial $nd$  system, respectively. All notation
are the same as in Fig.\ref{Fig: nd capture}.}
\end{figure}

For example, we concentrate on the detailed evaluation of the
diagram $a_4$ contribution. The energy and momentum of the incoming
particles are shown in Fig.\ref{Fig:a4}. We start by writing the
amplitude of the diagram in Fig.\ref{Fig:a4} using the Lagrangians
in Eqs.(\ref{Eq:1}) and (\ref{Eq:6}). Generally, before applying the
projection operators, we can write the contribution of the diagrams
in Fig.\ref{Fig:a4} in the cluster-configuration space up to
$\textrm{N}^n\textrm{LO}$ ($n\leq2$) as
\begin{eqnarray}\label{Eq:22}
\bar{S}^{(n)}_{4,unproj.}(E_i,k)=i\frac{e\,y^2}{16m_N}\,\int
\frac{d^4q}{(2\pi)^4}
\,t^{(n)^\dag}_{^3\!H}(q)\frac{1}{q_0-E_i+E_f-\frac{q^2}{2m_N}+i\varepsilon}\qquad\nonumber\\
\times\frac{1}{q_0-\frac{q^2}{2m_N}+i\varepsilon}
\,\frac{1}{E_i-q_0-\frac{k^2}{2m_N}-\frac{(\vec{k}+\vec{q})^2}{2m_N}+i\varepsilon}\mathcal{D}^{(n)}(E_i,q)\nonumber\\
\times\left(
                                                                         \begin{array}{cc}
                                                                           (k_0+k_1\tau_3)\sigma_k\sigma_s\sigma_rB_k & (k_0+k_1\tau_3)\sigma_k\tau_A\sigma_rB_k \\
                                                                           (k_0+k_1\tau_3)\sigma_k\sigma_s\tau_B B_k & (k_0+k_1\tau_3)\sigma_k\tau_A\tau_BB_k \\
                                                                         \end{array}
                                                                       \right),\qquad\!\!
\end{eqnarray}
where $k$ is the incoming momentum and
$E_i=\frac{3k^2}{4m_N}-\frac{\gamma_t^2}{m_N}$ denotes the energy of
the initial $nd$ system. $E_f$ represents the final state energy
which is given by $E_f=-B_t$ with $B_t=8.48$ MeV as the binding
energy of the triton. The $s$ ($A$) and $r$ ($B$) indices are the
spin (isospin) components of the incoming and outgoing dibaryons. To
solve the energy integration, we introduce the poles of
Eq.(\ref{Eq:22}) in the complex plane. It is obvious that we have
the three following pole:
\begin{eqnarray}\label{Eq:022}
q_0=\frac{q^2}{2m_N}-i\varepsilon,\qquad\qquad\qquad\quad\nonumber\\
q_0=E_i-E_f+\frac{q^2}{2m_N}-i\varepsilon,\qquad\;\nonumber\\
q_0=E_i-\frac{k^2}{2m_N}-\frac{(\vec{k}+\vec{q})^2}{2m_N}+i\varepsilon,
\end{eqnarray}
where they result from the denominator of the nucleon propagators.
With respect to the poles in Eq.(\ref{Eq:022}) and doing the
integration over energy and the solid angle, the
$\bar{S}^{(n)}_{4,unproj.}(E_i,k)$ is
\begin{eqnarray}\label{Eq:0022}
\bar{S}^{(n)}_{4,unproj.}(E_i,k)=\frac{e\,y^2}{32\pi^2}\frac{1}{E_f-E_i}\,\int^{\Lambda}_{0}dq\,q^2\,
t^{(n)^\dag}_{^3\!H}(q) \qquad\qquad\qquad\qquad\qquad\qquad\qquad\qquad\nonumber\\
\times\frac{1}{kq}
\bigg[\mathcal{D}^{(n)}(E_i,q)Q_0(\frac{m_NE_i-k^2-q^2}{kq})-\mathcal{D}^{(n)}(E_f,q)Q_0(\frac{m_NE_f-k^2-q^2}{kq})\bigg]\nonumber\\
\times\left(
                                                                         \begin{array}{cc}
                                                                           (k_0+k_1\tau_3)\sigma_k\sigma_s\sigma_rB_k & (k_0+k_1\tau_3)\sigma_k\tau_A\sigma_rB_k \\
                                                                           (k_0+k_1\tau_3)\sigma_k\sigma_s\tau_B B_k & (k_0+k_1\tau_3)\sigma_k\tau_A\tau_BB_k \\
                                                                         \end{array}
                                                                       \right).\qquad\qquad\qquad\qquad\qquad
\end{eqnarray}
with $Q_0(z)$ as the zeroth Legendre polynomial of the second kind.
In order to obtain the contribution of the diagram in
Fig.\ref{Fig:a4} for the $nd\rightarrow$ $^3H\gamma$
 reaction, we have to
project the initial $Nd$ system to the doublet and quartet cases
(corresponding to two possible M1 transitions) while the final state
should be $^2S_{\frac{1}{2}}$ because of the triton. The
contribution of the M1 transition with the initial quartet channel
($^4S_{\frac{3}{2}}$) is calculated by applying the projection
operators
\begin{eqnarray}\label{Eq:23}
\mathcal{P}_{d,rB}=\frac{1}{\sqrt{3}}\left(
                                         \begin{array}{cc}
                                           \sigma_r   & 0 \\
                                         0 & \tau_B \\
                                         \end{array}
                                       \right),
\end{eqnarray}
and
\begin{eqnarray}\label{Eq:24}
\mathcal{P}^s_{q,l}=\left(
                        \begin{array}{cc}
                         \delta^s_{l}-\frac{1}{3}\sigma^s\sigma_l   & 0 \\
                                         0 & 0\\
                        \end{array}
                      \right),
\end{eqnarray}
with $l$ as the spin component of the deuteron in the quartet
channel, from left and right in Eq.(\ref{Eq:0022}), respectively.
So, we gain
\begin{eqnarray}\label{Eq:26}
\bar{S}^{(n)}_{4,q}(E_i,k)=\frac{e\,y^2}{24\sqrt{3}\pi^2}\frac{1}{E_f-E_i}\,\int^{\Lambda}_{0}dq\,q^2\,
t^{(n)^\dag}_{^3\!H}(q)\qquad\qquad\qquad\qquad\qquad\qquad\qquad\qquad\quad\;\nonumber\\
\times\frac{1}{kq}
\bigg[\mathcal{D}^{(n)}(E_i,q)Q_0(\frac{m_NE_i-k^2-q^2}{kq})
-\mathcal{D}^{(n)}(E_f,q)Q_0(\frac{m_NE_f-k^2-q^2}{kq})\bigg]
\nonumber\\ \times \left(
                                                                         \begin{array}{cc}
                                                                           k_0+k_1\tau_3 & 0 \\
                                                                           0 & 0 \\
                                                                         \end{array}
                                                                       \right)
t^\dag\big[2\,i\vec{\varepsilon}_d\cdot\vec{\varepsilon}^\ast_\gamma
\times\vec{\tilde{q}}-\vec{\sigma}\times\vec{\varepsilon}_d\cdot\vec{\varepsilon}^\ast_\gamma\times\vec{\tilde{q}}\big]N.\qquad\qquad\qquad\qquad\;\,
\end{eqnarray}

Taking into account the projection operators
$\mathcal{P}^{sA}_{d}=\big(\mathcal{P}_{d,sA}\big)^\dag$ and
$\mathcal{P}_{d,rB}$ for the incoming and outgoing channels,
respectively, the contribution of the diagram in Fig.\ref{Fig:a4}
for the initial doublet channel is given by
\begin{eqnarray}\label{Eq:25}
\bar{S}^{(n)}_{4,d}(E_i,k)=\frac{e\,y^2}{96\pi^2}\frac{1}{E_f-E_i}\,\int^{\Lambda}_{0}dq\,q^2\,
t^{(n)^\dag}_{^3\!H}(q)\qquad\qquad\qquad\qquad\qquad\qquad\qquad\qquad\qquad\quad\nonumber\\
\times\frac{1}{kq}
\bigg[\mathcal{D}^{(n)}(E_i,q)Q_0(\frac{m_NE_i-k^2-q^2}{kq})
-\mathcal{D}^{(n)}(E_f,q)Q_0(\frac{m_NE_f-k^2-q^2}{kq})\bigg]
\nonumber\\ \times \left(
                                                                         \begin{array}{cc}
                                                                           k_0+k_1\tau_3 & -3(k_0+k_1\tau_3) \\
                                                                           3(3k_0-k_1\tau_3) & -(3k_0-k_1\tau_3) \\
                                                                         \end{array}
                                                                       \right)
t^\dag\big[i\vec{\varepsilon}_d\cdot\vec{\varepsilon}^\ast_\gamma
\times\vec{\tilde{q}}+\vec{\sigma}\times\vec{\varepsilon}_d\cdot\vec{\varepsilon}^\ast_\gamma\times\vec{\tilde{q}}\big]N.\qquad\!\!
\end{eqnarray}
The results of Eqs.(\ref{Eq:26}) and (\ref{Eq:25}) are calculated
using $\vec{B}=-i\vec{\tilde{q}}\times\vec{\varepsilon}^\ast_\gamma$
and the sum over the repeated indices.

Finally, the total contribution of the diagram in Fig.\ref{Fig:a4}
can be written as
\begin{eqnarray}\label{Eq:27}
W^{(n)}_{4}=\bar{S}^{(n)}_{4,d}+\bar{S}^{(n)}_{4,q}=t^\dag\big[S^{(n)}_{4,d}Y_d+S^{(n)}_{4,q}Y_q]N.
\end{eqnarray}
where $t^\dag S^{(n)}_{4,x}Y_x N=\bar{S}^{(n)}_{4,x}$ with $x=d,q$.
By ignoring the normalization factor of the incoming deuteron,
Eq.(\ref{Eq:27}) is as we expected.

One can evaluate the contribution of all diagrams in Fig.\ref{Fig:
nd capture} using the same procedure as for the $a_4$ diagram. After
applying the integration over energy and solid angle, the
contribution of all M1 diagrams in the $\mathcal{M}^{(n)}_x$
function (Eq.(\ref{Eq:8})), before multiplying the deuteron wave
function normalization factor, is given by
\begin{eqnarray}\label{Eq:10}
M^{(n)}_x(E_i,k)=S_{0,x}^{(n)}(E_i,k)+S_{x}^{(n)}(E_i,k)\qquad\qquad\qquad\qquad\qquad\qquad\quad \nonumber\\
-\frac{1}{2\pi^2}\int_0^{\Lambda}
dq\,q^2\,S_{x}^{(n)}(E_i,q)\,\mathcal{D}^{(n)}(E_i,q)\,t_{x}^{(n)}(E_i;k,q)
\end{eqnarray}
where
\begin{eqnarray}\label{Eq:11}
S_{x}^{(n)}(E_i,k)=\sum_{i=1}^{5}S_{i,x}^{(n)}(E_i,k).
\end{eqnarray}
In the above, $x$ can be $"d"$ or $"q"$ for doublet and quartet
channels, respectively. The $2\times2$ matrix function
$S_{i,x}^{(n)}$ with $i=0,\cdot\cdot\cdot,5$ represents the
contribution of the $"a_i"$ diagram in Fig.\ref{Fig: nd capture} for
the initial $x$ channel up to $\textrm{N}^n\textrm{LO}$ ($n\leq2$).

For the initial doublet ($S=\frac{1}{2}$) state, in the
cluster-configuration space, we obtain
\begin{eqnarray*}\label{Eq:1200000}
S_{0,d}^{(n)}(E_i,k)=\frac{e}{6m_N}\frac{1}{E_f-E_i}\,
t^{(n)^\dag}_{^3\!H}(k)\left(
                                                                         \begin{array}{cc}
                                                                           -(k_0+k_1\tau_3) & 0 \\
                                                                           0 & 3k_0-k_1\tau_3 \\
                                                                         \end{array}
                                                                       \right)
 ,\qquad\qquad\qquad\qquad
\end{eqnarray*}
\begin{eqnarray*}\label{Eq:120}
S_{1,d}^{(n)}(E_i,k)=\frac{e\,y^2}{32\pi}\frac{1}{E_f-E_i}\,
t^{(n)^\dag}_{^3\!H}(k)\mathcal{D}^{(n)}(E_f,k)\qquad\qquad\qquad\qquad\qquad\qquad\qquad\qquad\quad\,\nonumber\\
\times
\Bigg[\sqrt{\frac{3}{4}k^2-m_NE_i}-\sqrt{\frac{3}{4}k^2-m_NE_f}\Bigg]\left(
                                                                         \begin{array}{cc}
                                                                           2k_0 & k_1\tau_3 \\
                                                                           k_1\tau_3 & 0 \\
                                                                         \end{array}
                                                                       \right) ,
\end{eqnarray*}
\begin{eqnarray*}\label{Eq:1200}
S_{2,d}^{(n)}(E_i,k)=\frac{e\,y^2}{96\pi^2}\frac{1}{E_f-E_i}\,\int^{\Lambda}_{0}dq\,q^2
t^{(n)^\dag}_{^3\!H}(q)\mathcal{D}^{(n)}(E_f,q)\frac{1}{kq}
\bigg[Q_0(\frac{m_NE_i-k^2-q^2}{kq}) \qquad\nonumber\\
-Q_0(\frac{m_NE_f-k^2-q^2}{kq})\bigg]\left(
                                                                         \begin{array}{cc}
                                                                           -5k_0+5k_1\tau_3 & 3k_0+k_1\tau_3 \\
                                                                           3k_0+k_1\tau_3 & 3k_0+5k_1\tau_3 \\
                                                                         \end{array}
                                                                       \right)
,
\end{eqnarray*}
\begin{eqnarray*}\label{Eq:12000}
S_{3,d}^{(n)}(E_i,k)=\frac{e}{3m_N\rho_d}\,
t^{(n)^\dag}_{^3\!H}(k)\mathcal{D}^{(n)}(E_f,k)\left(
                                                                         \begin{array}{cc}
                                                                           4 L_2 & \sqrt{\frac{\rho_d}{\,r_0}} L_1\tau_3 \\
                                                                           \sqrt{\frac{\rho_d}{\,r_0}} L_1\tau_3 & 0 \\
                                                                         \end{array}
                                                                       \right)
                                                                       ,\qquad\qquad\qquad\qquad
\end{eqnarray*}
\begin{eqnarray*}\label{Eq:120000}
S_{4,d}^{(n)}(E_i,k)=\frac{e\,y^2}{96\pi^2}\frac{1}{E_f-E_i}\,\int^{\Lambda}_{0}dq\,q^2\,
t^{(n)^\dag}_{^3\!H}(q)\frac{1}{kq}
\bigg[\mathcal{D}^{(n)}(E_i,q)Q_0(\frac{m_NE_i-k^2-q^2}{kq})\qquad\nonumber\\
-\mathcal{D}^{(n)}(E_f,q)Q_0(\frac{m_NE_f-k^2-q^2}{kq})\bigg] \left(
                                                                         \begin{array}{cc}
                                                                           k_0+k_1\tau_3 & -3(k_0+k_1\tau_3) \\
                                                                           3(3k_0-k_1\tau_3) & -(3k_0-k_1\tau_3) \\
                                                                         \end{array}
                                                                       \right) ,
\end{eqnarray*}
\begin{eqnarray}\label{Eq:12}
S_{5,d}^{(n)}(E_i,k)=\frac{e\,y^2}{24\pi^2}\frac{1}{E_f-E_i}\,\mathcal{H}(E_i,\Lambda)\,\int^{\Lambda}_{0}dq\,q^2\,
t^{(n)^\dag}_{^3\!H}(q) \qquad\qquad\qquad\qquad\qquad\qquad\qquad\!\!\!\nonumber\\
\times \bigg[\mathcal{D}^{(n)}(E_i,q)-\mathcal{D}^{(n)}(E_f,q)\bigg]
\left(
                                                                         \begin{array}{cc}
                                                                           k_0+k_1\tau_3 & -(k_0+k_1\tau_3) \\
                                                                           3k_0-k_1\tau_3 & -(3k_0-k_1\tau_3) \\
                                                                         \end{array}
                                                                       \right).
\end{eqnarray}
Also, in the incoming quartet channel ($S=\frac{3}{2}$), we have
\begin{eqnarray*}\label{Eq:130}
S_{0,q}^{(n)}(E_i,k)=\frac{e}{3\sqrt{3}m_N}\frac{1}{E_f-E_i}\,
t^{(n)^\dag}_{^3\!H}(k)\left(
                                                                         \begin{array}{cc}
                                                                           k_0+k_1\tau_3 & 0 \\
                                                                           0 & 0 \\
                                                                         \end{array}
                                                                       \right) ,\qquad\qquad\qquad\qquad\qquad\quad
\end{eqnarray*}
\begin{eqnarray*}\label{Eq:1300}
S_{1,q}^{(n)}(E_i,k)=\frac{e\,y^2}{32\sqrt{3}\pi}\frac{1}{E_f-E_i}\,
t^{(n)^\dag}_{^3\!H}(k)\mathcal{D}^{(n)}(E_f,k)\qquad\qquad\qquad\qquad\qquad\qquad\qquad\;\,\;\nonumber\\
\times
\Bigg[\sqrt{\frac{3}{4}k^2-m_NE_i}-\sqrt{\frac{3}{4}k^2-m_NE_f}\Bigg]\left(
                                                                         \begin{array}{cc}
                                                                           2k_0 & 0 \\
                                                                           k_1\tau_3 & 0 \\
                                                                         \end{array}
                                                                       \right) ,
\end{eqnarray*}
\begin{eqnarray*}\label{Eq:13000}
S_{2,q}^{(n)}(E_i,k)=\frac{e\,y^2}{48\sqrt{3}\pi^2}\frac{1}{E_f-E_i}\int^{\Lambda}_{0}dqq^2
t^{(n)^\dag}_{^3\!H}(q)\mathcal{D}^{(n)}(E_f,q)\frac{1}{kq}
\bigg[Q_0(\frac{m_NE_i-k^2-q^2}{kq}) \nonumber\\
-Q_0(\frac{m_NE_f-k^2-q^2}{kq})\bigg]\left(
                                                                         \begin{array}{cc}
                                                                           -k_0+k_1\tau_3 & 0 \\
                                                                           -3k_0-k_1\tau_3 & 0 \\
                                                                         \end{array}
                                                                       \right) ,
\end{eqnarray*}
\begin{eqnarray*}\label{Eq:13000}
S_{3,q}^{(n)}(E_i,k)=\frac{e}{3\sqrt{3}m_N\rho_d}\,
t^{(n)^\dag}_{^3\!H}(k)\mathcal{D}^{(n)}(E_f,k)\left(
                                                                         \begin{array}{cc}
                                                                          -2 L_2 & 0 \\
                                                                           \sqrt{\frac{\rho_d}{\,r_0}} L_1\tau_3 & 0 \\
                                                                         \end{array}
                                                                       \right) ,\qquad\qquad\qquad\qquad\;\;
\end{eqnarray*}
\begin{eqnarray*}\label{Eq:130000}
S_{4,q}^{(n)}(E_i,k)=\frac{e\,y^2}{24\sqrt{3}\pi^2}\frac{1}{E_f-E_i}\int^{\Lambda}_{0}dqq^2
t^{(n)^\dag}_{^3\!H}(q)\frac{1}{kq}
\bigg[\mathcal{D}^{(n)}(E_i,q)Q_0(\frac{m_NE_i-k^2-q^2}{kq}) \!\nonumber\\
-\mathcal{D}^{(n)}(E_f,q)Q_0(\frac{m_NE_f-k^2-q^2}{kq})\bigg] \left(
                                                                         \begin{array}{cc}
                                                                           k_0+k_1\tau_3 & 0 \\
                                                                           0 & 0 \\
                                                                         \end{array}
                                                                       \right) ,
\end{eqnarray*}
\begin{eqnarray}\label{Eq:13}
S_{5,q}^{(n)}(E_i,k)=0
.\qquad\qquad\qquad\qquad\qquad\qquad\qquad\qquad\qquad\qquad\qquad\qquad\qquad\qquad\;\;
\end{eqnarray}

The results of $M^{(n)}_{i,x}$ are obtained after applying the
appropriate projection operators for initial and final states. We
note that the $S_{5,q}^{(n)}$ must be zero since in the quartet
($S$=$\frac{3}{2}$) channel all spins are aligned and there is no
three-body interaction in this channel because the Pauli principle
forbids the three nucleons to be at the same point in space.

Low-energy observables of the $nd\rightarrow$ $^3H\gamma$ process
are cutoff-independent by the introduction of $H_0$ and $H_2$ up to
$\textrm{N}^2\textrm{LO}$  (see Table \ref{tab:cutoff var}). Namely,
they are renormalized and therefore no new three-body forces are
needed up to $\textrm{N}^2\textrm{LO}$. The same argument can be
applied equally with three-body currents \cite{Griesshammer}, so no
three-body currents are included in the present calculation.

We stress that the $M^{(n)}_{x}$ amplitude is a $2\times2$ matrix
which is written in the cluster-configuration space and so the
contributions of both initial $nd_t$ and $nd_s$ systems are taken
into account. Thus, the physical amplitude of the $nd\rightarrow$
$^3H\gamma$ process is given by
\begin{eqnarray}\label{Eq:14}
\mathcal{M}^{(n)}_x(E;k,p)=M^{(n)}_x(E;k,p)\cdot\left(
                                             \begin{array}{c}
                                               \sqrt{\mathcal{Z}^{(n)}_t} \\
                                               0 \\
                                             \end{array}
                                           \right)
\end{eqnarray}
where $\mathcal{Z}^{(n)}_t$ indicates the normalization factor of
the incoming deuteron wave function at $\textrm{N}^n\textrm{LO}$,
\begin{eqnarray}\label{Eq:15}
\mathcal{Z}^{(n)}_t=\Bigg(\frac{\partial}{\partial
q_0}\frac{1}{D_{t}^{(n)}(q_0,q)} \Bigg
|_{q_0=-\frac{\gamma_t^2}{m_N},q=0}\Bigg)^{-1}.
\end{eqnarray}
We note that $\tau_3=-1$ must be applied for the $nd\rightarrow$
$^3H\gamma$ process.

\section{Cross section of $nd\rightarrow$ $^3H\gamma$ process}\label{Sec:cross section}
In the following, we use the $\mathcal{W}^{(n)}$ amplitude for
calculating the total cross section of the $nd\rightarrow$
$^3H\gamma$. In order to proceed to calculate the cross section, we
use the following spin sums:
\begin{eqnarray}\label{Eq:15}
\sum_{spin/pol}(t^\dag Y_d N)(N^\dag Y_q^\dag t)=0,
\nonumber\\
\frac{1}{6}\sum_{spin/pol}\Big|t^\dag Y_d N\Big|^2=\frac{2}{9},
\quad\nonumber\\
\frac{1}{6}\sum_{spin/pol}\Big|t^\dag Y_q
N\Big|^2=\frac{4}{9},\quad\,
\end{eqnarray}
where the factor $\frac{1}{6}$ comes from the average over initial
state polarizations. The above calculations are done in the Coulomb
gauge $\vec{\tilde{q}}\cdot\vec{\varepsilon}_\gamma$ and the results
are given using
$\varepsilon^{i^\ast}_d\varepsilon^{j}_d=\varepsilon^{j^\ast}_d\varepsilon^{i}_d=\delta_{ij}$,
$\vec{\tilde{q}}=(0,0,1)$ and
$\vec{\varepsilon}^{\pm}_\gamma=\frac{1}{\sqrt{2}}(1,\mp i,0)$,
where the upper and lower signs denote the photon with the right and
left helicity, respectively.

From Eq.(\ref{Eq:15}), the total cross section of the neutron
radiative capture by a deuteron can be written as
\begin{eqnarray}\label{Eq:16}
\sigma^{(n)}_{tot}=\frac{(E_i-E_f)^3}{v}\,\frac{\big|\mathcal{M}^{(n)}_d\big|^2+2\,\big|\mathcal{M}^{(n)}_q\big|^2}{27}
\end{eqnarray}
where $"^{(n)}"$ superscript denotes $\textrm{N}^n\textrm{LO}$
results and $v$ is the incident neutron velocity in the c.m. frame.

\section{Numerical implementation}\label{sec:numerical}
In the computation of the M1 amplitude of the diagrams in
Fig.\ref{Fig: nd capture}, we need to obtain the triton wave
function and the half-off-shell $Nd$ scattering amplitude at leading
(LO), next-to-leading, and next-to-next-to-leading orders. The
half-off-shell neutron-deuteron scattering is obtained order by
order by solving numerically the Faddeev equations which are
introduced in Appendix A for both initial doublet and quartet
channels. We solve them by the Hetherington-Schick method
\cite{Hetherington-Schick,Cahill-Sloan,Aaron-Amado} in a
$\textit{Mathematica}$ code with a specific cutoff momentum
$\Lambda$. We also obtain the triton wave function at each order by
solving the homogenous part of the Faddeev equations of $Nd$
scattering in the doublet channel with the same cutoff and then
normalize it by the method which is introduced in Appendix
\ref{Appendix B}.

Using the order-by-order results of $t^{(n)}_{^3\!H}$ and
$t^{(n)}_{x}$ ($x=d$ and $ q $), we can be able to solve the
integrations in Eqs.(\ref{Eq:10}), (\ref{Eq:12}) and (\ref{Eq:13})
to obtain the M1 amplitude of $nd\rightarrow$ $^3H\gamma$. We solve
these integrations numerically using the Gaussian quadrature weights
and also the same cutoff momentum $\Lambda$ as before.

As we see from Eq.(\ref{Eq:05}), the parameter $H_0(\Lambda)$ and
$H_2(\Lambda)$ must be determined order by order for the cutoff
$\Lambda$. At each order, we obtain the value of $H_0$ by
constructing the exact triton scattering length, $a_3=0.65$ fm. The
$H_2$ parameter which enters at $\textrm{N}^2\textrm{LO}$ is
determined for an arbitrary cutoff value by matching the triton
binding energy to the experimental value, $B_t^{exp}=8.48$ MeV.

\section{Results}\label{sec:results}

\begin{figure}
\includegraphics*[width=16cm]{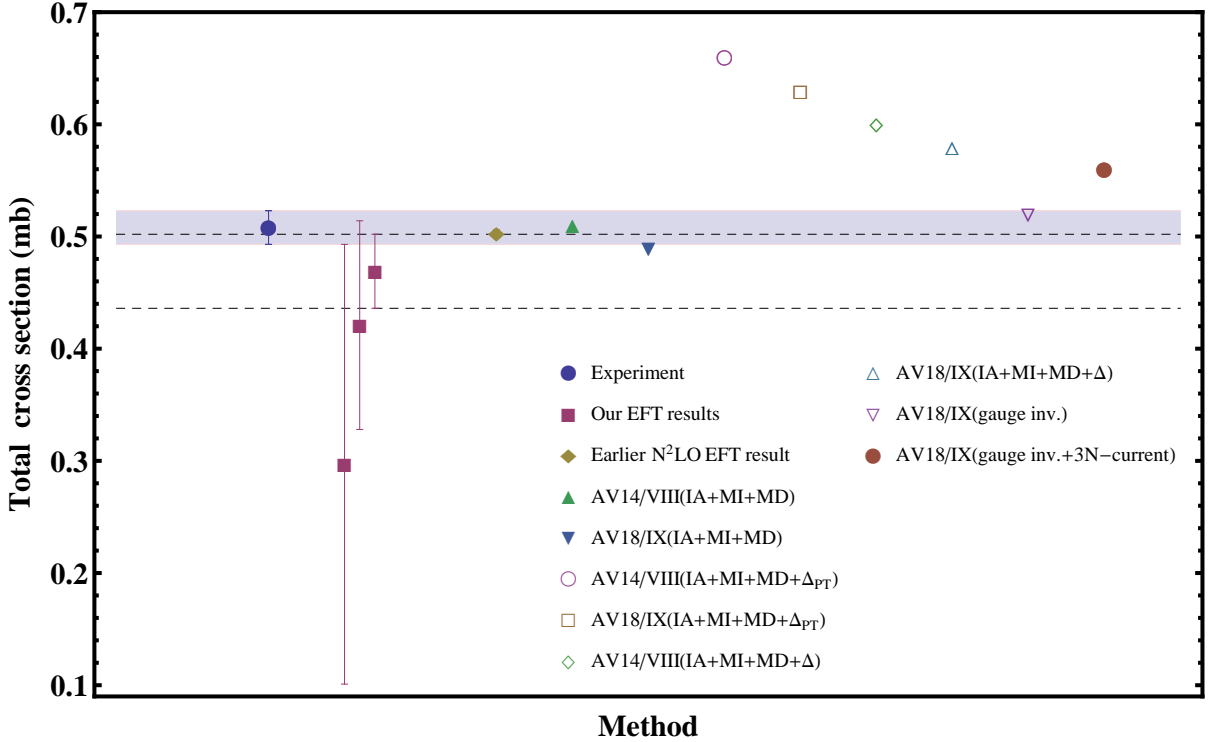}\centering
\caption{\label{Fig:plot}Comparison between different theoretical
results for the total cross section of the $nd\rightarrow$
$^3H\gamma$ process. The points from left to right denote the
results computed by experiment \cite{Jurney et al}, our LO
EFT($\pi\!\!\!/$), our NLO EFT($\pi\!\!\!/$), our
$\textrm{N}^2\textrm{LO}$ EFT($\pi\!\!\!/$), earlier
$\textrm{N}^2\textrm{LO}$ EFT($\pi\!\!\!/$)
\cite{sadeghi-bayegan-grieshammer}, AV14/VIII(IA+MI+MD)
\cite{viviani et al}, AV18/IX(IA+MI+MD) \cite{viviani et al},
AV14/VIII(IA+MI+MD+$\Delta_{PT}$) \cite{viviani et al},
AV18/IX(IA+MI+MD+$\Delta_{PT}$) \cite{viviani et al},
AV14/VIII(IA+MI+MD+$\Delta$) \cite{viviani et al},
AV18/IX(IA+MI+MD+$\Delta$) \cite{viviani et al}, AV18/IX(gauge inv.)
\cite{Marcucci et al}, and AV18/IX(gauge inv.+$3N$-current)
\cite{Marcucci et al} methods, respectively. The thin band indicates
the error band of the experimental result of the cross section. Two
horizontal dashed lines determine the upper and lower limits due to
our systematic EFT($\pi\!\!\!/$) error at
$\textrm{N}^2\textrm{LO}$.}
\end{figure}

In this work, we have concentrated on the evaluation of the cross
section of the $nd\rightarrow$ $^3H\gamma$ process up to
$\textrm{N}^2\textrm{LO}$. Our EFT($\pi\!\!\!/$) results for the
amplitudes and cross sections of the $nd\rightarrow$ $^3H\gamma$
process at thermal energy, $2.5\times10^{-8} $ MeV are shown in
Table \ref{tab:results}. We compare schematically our
EFT($\pi\!\!\!/$) results at thermal energy for the cross section
with the previous model-dependent theoretical calculations and the
experimental data in Fig.\ref{Fig:plot}.

\begin{table}[h]\centering
\caption{Our EFT($\pi\!\!\!/$) results for the amplitudes and cross
sections of the $nd\rightarrow$ $^3H\gamma$ process at thermal
energy, $2.5\times10^{-8} $ MeV. $n$ denotes our results up to
$\textrm{N}^n\textrm{LO}$. $\mathcal{M}^{(n)}_x$ and
$\sigma^{(n)}_x$ are the amplitude and cross section of the
$nd\rightarrow$ $^3H\gamma$ process for the incoming $x$ ($x=d,q$)
channel up to $\textrm{N}^n\textrm{LO}$, respectively.
$\sigma^{(n)}_{tot}$ is the total cross section up to
$\textrm{N}^n\textrm{LO}$. The deviations which have been added to
our EFT($\pi\!\!\!/$) results of the total cross section indicates
the systematic EFT($\pi\!\!\!/$) errors at each order. The results
of the amplitudes and cross sections are presented in
$10^{-7}\;\textrm{MeV}^{-\frac{5}{2}}$ and mb units, respectively.}
\label{tab:results}       
\begin{tabular}{cccccc}
\hline\hline\noalign{\smallskip}$\qquad n\qquad$ & $\quad
\sqrt{|\mathcal{M}^{(n)}_q|^2} \quad $ &
 $\qquad\sqrt{|\mathcal{M}^{(n)}_d|^2}\qquad$& $\qquad\sigma^{(n)}_q\qquad$ &
 $\qquad\sigma^{(n)}_d\qquad$
& $\qquad\sigma^{(n)}_{tot}\qquad$  \\
\noalign{\smallskip} \hline\noalign{\smallskip}

$0$  &$4.88$ & $3.65$ & 0.232 & 0.065 & 0.297 $\pm$ 0.196  \\

$1$ & $5.20$ &$5.68$ & 0.264 & 0.157 & 0.421 $\pm$ 0.093  \\

$2$  & $5.30$ & $6.33$ & 0.273 & 0.196 & 0.469 $\pm$ 0.033\\


 \noalign{\smallskip}\hline\hline
\end{tabular}
\end{table}

We use the power counting introduced by Bedaque $\textit{et al.}$ in
\cite{Bedaque-H-vK,Bedaque-H-vK-2,Bedaque-R-H}. The
EFT($\pi\!\!\!/$) expansion parameter is
$\frac{Q}{\bar{\Lambda}}\sim\frac{1}{3}$, where $Q$ and
$\bar{\Lambda}$ are the small and large parameters, so the NLO and
$\textrm{N}^2\textrm{LO}$ diagrams enter 33$\%$ and 11$\%$
corrections to the leading- and next-to-leading-order amplitudes,
respectively. Also, with respect to our power counting, the error of
$\textrm{N}^2\textrm{LO}$ amplitude must be less than 3.7$\%$ of the
exact value. The cross section is proportional to the square of the
amplitude. It is obvious that if the systematic EFT($\pi\!\!\!/$)
error in the amplitude is $"\alpha\%"$, as an example, the cross
section has a maximum error $\thicksim2\alpha\%$ in the
EFT($\pi\!\!\!/$) approach. So, we expect to have a maximum error of
7$\%$ at $\textrm{N}^2\textrm{LO}$ for the cross section.

Our results in Table \ref{tab:results} show the convergence in our
power counting from LO to $\textrm{N}^2\textrm{LO}$. At NLO, 0.124
mb adds to the leading-order value and at $\textrm{N}^2\textrm{LO}$
0.048 mb to the next-to-leading order. Our EFT result for the total
cross section of neutron radiative capture by a deuteron at
$\textrm{N}^2\textrm{LO}$, $\sigma_{tot}^{(2)}=0.469$ mb, has an
error of 7$\%$ compare with the experimental value,
$\sigma_{tot}^{exp}=0.508\pm0.015$ mb. We stress that the
contribution of the E2 transition has not been included in our
calculation for the amplitude of the $nd\rightarrow$ $^3H\gamma$
reaction. The E2 transition is suppressed by two powers of the
initial nucleon momentum or photon energy compared to the dominant
M1 transition. Therefore, this effect numerically has a contribution
of $(\frac{(E_i-E_f) \;\textrm{or}\; k}{\Lambda})^2\sim 0.25\%$
correction in the quartet-initial-channel amplitude of
$nd\rightarrow$ $^3H\gamma$ and so $\sim 0.5\%$ in total cross
section at threshold regime. Also, with respect to the power
counting as discussed above, we expect a maximum error
$\thicksim7\%$ in $\textrm{N}^2\textrm{LO}$ EFT($\pi\!\!\!/$)
results of the cross section. Thus the 7$\%$ error in our
$\textrm{N}^2\textrm{LO}$ results is acceptable. We believe that the
higher-order corrections make this discrepancy narrow.
\begin{table}[h]\centering
\caption{The cutoff variation of our EFT($\pi\!\!\!/$) results for
the total cross section between $\Lambda=200$ and $\Lambda=900$ MeV.
$n=0,1,2$ denote the LO, NLO and $\textrm{N}^2\textrm{LO}$ results,
respectively.}
\label{tab:cutoff var}       
\begin{tabular}{cc}
\hline\hline\noalign{\smallskip} $\qquad n\qquad$ &
Abs[$1-\frac{\sigma^{(n)}_{tot}(\Lambda=200\,\textrm{MeV})}{\sigma^{(n)}_{tot}(\Lambda=900\,\textrm{MeV})}$]  \\
\noalign{\smallskip} \hline\noalign{\smallskip}

$0$  & 0.098756  \\

$1$ & 0.045714 \\

$2$  & 0.004006 \\


 \noalign{\smallskip}\hline\hline
\end{tabular}
\end{table}

According to Table \ref{tab:cutoff var}, we have computed the cutoff
variation of our EFT($\pi\!\!\!/$) results for the total cross
section within a natural range of $\Lambda=200$ to $\Lambda=900$ MeV
at LO, NLO, and $\textrm{N}^2\textrm{LO}$. The range of cutoff
variation should be a few times the pion mass because, here, the
existence of a definite $\Lambda\rightarrow\infty$ limit in an EFT
calculation does not guarantee that the results found in that limit
are rigorous consequences of the EFT \cite{Epelbanm,Ji-Phillips}.
Our results in Table \ref{tab:cutoff var} indicate that the M1
amplitudes and the cross section of $nd\rightarrow$ $^3H\gamma$ are
cutoff independent and properly renormalized.

The differences of our results and the previous EFT($\pi\!\!\!/$)
calculation of total cross section at thermal energy
\cite{sadeghi-bayegan-grieshammer} are due to the ignored diagrams
and the $^3S_1\rightarrow^3$$S_1$ M1 transition effects.

\begin{table}[h]\centering
\caption{The $^3S_1\rightarrow^3$$S_1$ M1 transition effect in the
total cross section at LO ($n=0$), NLO ($n=1$) and
$\textrm{N}^2\textrm{LO}$ ($n=2$).}
\label{tab:L2 effect}       
\begin{tabular}{|c||c|c|c|}
\noalign{\smallskip}\hline $\; n\;$ &
$\qquad\sigma_{tot}^{(n)}\:(L_2=0)\qquad$ & $\quad\sigma_{tot}^{(n)}\:(L_2=-0.4\;\textrm{fm})\quad$ & $\qquad$ difference$\qquad$ \\
\hline\noalign{\smallskip} \hline

$0$  & $0.297\pm0.196$ & $0.297\pm0.196$ & 0 \\

$1$ & $0.472\pm0.104$ &  $0.421\pm0.093$&  $0.051$\\

$2$  & $0.553\pm0.041$ & $0.469\pm0.033$ & $0.084$\\


\hline \noalign{\smallskip}
\end{tabular}
\end{table}

The $L_2$ coefficient corresponding to the contribution of the
$^3S_1\rightarrow^3$$S_1$ M1 transition is small compared with $L_1$
which comes from the $^1S_0\rightarrow^3$$S_1$ M1 transition
\cite{Ando-Hyun}. So, we expect that the $^3S_1\rightarrow^3$$S_1$
M1 transition has a small (and negligible) effect at NLO results but
at $\textrm{N}^2\textrm{LO}$ the $^3S_1\rightarrow^3$$S_1$ M1
transition could have a significant effect. Our results for the
total cross section with and without the $L_2$ coefficient effect
which are summarized in Table \ref{tab:L2 effect} are as we
expected.

\begin{table}[h]\centering
\caption{The investigation of the ignored contributions in the
previous EFT($\pi\!\!\!/$) calculation
\cite{sadeghi-bayegan-grieshammer} at each order.
$\bar{\mathcal{M}}^{(n)}_{x}$ indicates the total
$\textrm{N}^n\textrm{LO}$ amplitude of M1 $nd\rightarrow$
$^3H\gamma$ without the contribution of the time ordering that
corresponds to the photon emitted during the nucleon exchange in the
first diagram of the second line in Fig.\ref{Fig: nd capture}.
$\mathcal{M}^{(n)}_{013,x}$ denotes the sum of the amplitudes of the
diagrams "$a_0$", "$a_1$", and "$a_3$" for the incoming $x$ channel.
Also, $\mathcal{M}^{(n)}_{2,x}$ is only the contribution of the
"$a_2$" diagram in the first line of Fig.\ref{Fig: nd capture} for
the initial $x$ channel, respectively. $q$ and $d$ in the first
column indicate the initial quartet ($^4$$S_{\frac{3}{2}}$) and
doublet ($^2$$S_{\frac{1}{2}}$) channels, respectively. The results
of the amplitudes are presented in units of
$10^{-7}\;\textrm{MeV}^{-\frac{5}{2}}$. }
\label{tab:ignored diagrams effects}       
\begin{tabular}{|c|c||c|c|c|c|}
\hline $\;\; x\;\;$ & $\;\;n\;\;$ &
 $\qquad\sqrt{|\mathcal{M}^{(n)}_{x}|^2}\qquad$ & $ Abs\Big[\sqrt{|\mathcal{M}^{(n)}_{x}|^2}-\sqrt{|\bar{\mathcal{M}}^{(n)}_{x}|^2}\Big] $ & $\quad \sqrt{|\mathcal{M}^{(n)}_{013,x}|^2} \quad$ & $\quad\sqrt{|\mathcal{M}^{(n)}_{2,x}|^2}\quad$  \\
\hline \hline

 &$0$  &  $4.88\pm1.61$ & $0.10$ & $5.27$  & $0.21$  \\

$q$ &$1$  & $5.20\pm0.57$ & $0.47$ & $6.05$  & $0.42$  \\

 &$2$  & $5.30\pm0.19$ & $0.90$ & $6.26$ & $0.58$  \\
\hline
 &$0$  & $3.65\pm1.20$ & $2.11$ &$3.79$ & $1.16$  \\

$d$ &$1$  & $5.68\pm0.63$ & $2.94$ & $4.37$ & $2.41$ \\

 &$2$  & $6.33\pm0.23$ & $3.47$ & $4.55$ & $3.39$ \\


 \hline
\end{tabular}
\end{table}

The effects of the diagrams in Fig.\ref{Fig: nd capture} which have
been neglected in the previous EFT($\pi\!\!\!/$) calculation
\cite{sadeghi-bayegan-grieshammer} have been investigated in Table
\ref{tab:ignored diagrams effects}. The results in the third column
of Table \ref{tab:ignored diagrams effects} are the total doublet
and quartet amplitudes of the M1 $nd\rightarrow$ $^3H\gamma$
transition at LO, NLO and $\textrm{N}^n\textrm{LO}$. In the fourth
column, we present the computed values of the contribution which is
only corresponding to the nucleon pole before photon creation in the
first diagram in the second line of Fig.\ref{Fig: nd capture} at
each order. The fifth and sixth columns of Table \ref{tab:ignored
diagrams effects} represent only the evaluated values for the
amplitudes of the "$a_0+a_1+a_3$" and "$a_2$" diagrams in the first
line of Fig.\ref{Fig: nd capture}, respectively, for both doublet
and quartet channels.

The lack of a correct calculation of the first diagram of the second
line in Fig.\ref{Fig: nd capture} creates the significant errors as
indicated in the fourth column of Table IV at each order. The
results shown in the fifth column of Table \ref{tab:ignored diagrams
effects} indicate that the diagrams with radiation from external
nucleon leg, external deuteron leg, and on-shell two-body bubble in
the first line of Fig.\ref{Fig: nd capture} are LO effects and so,
one expects that these diagrams have a very important effect in the
final results of the amplitude of the M1 $nd\rightarrow$ $^3H\gamma$
transition and could not be ignored. But the last column depicts
that the diagram "$a_2$" in the first line of Fig.\ref{Fig: nd
capture} has a small effect at LO especially in the quartet channel
as estimated in the previous EFT($\pi\!\!\!/$) calculation
\cite{sadeghi-bayegan-grieshammer}. Finally, we emphasize that our
results have been evaluated using the properly normalized triton
wave function.

\section{Conclusion and Outlook} \label{sec:conclusion}
In the present parity-conserving EFT($\pi\!\!\!/$) calculation, we
have calculated the amplitudes and cross section for $nd\rightarrow$
$^3H\gamma$ fully in the cluster-configuration space up to
$\textrm{N}^2\textrm{LO}$. We have considered one- and two-body
currents. No three-body currents are needed to renormalize the
observables in this work up to $\textrm{N}^2\textrm{LO}$. The M1 is
the dominant transition at the low energies. We have included the
contribution of the possible diagrams which have not been included
in the previous EFT($\pi\!\!\!/$) calculations and used the properly
normalized triton wave function. We have also considered the effects
of the $^3S_1\rightarrow^3$$S_1$ M1 transition ($L_2$ coefficient in
Eq.(\ref{Eq:6})) in the $nd\rightarrow$ $^3H\gamma$ amplitudes
together with other transitions which are included in the previous
EFT($\pi\!\!\!/$) calculations.

The $\textrm{N}^2\textrm{LO}$ EFT($\pi\!\!\!/$) total cross section
is determined to be $\sigma^{(2)}_{tot}=0.469\pm0.033$ mb. The
reliable calculation of the doublet and quartet amplitudes can be
used in the calculation of parity-violating observables in the
$nd\rightarrow$ $^3H\gamma$ process. The $\textrm{N}^2\textrm{LO}$
EFT($\pi\!\!\!/$) total cross section $\sigma_{tot}^{(2)}$ is within
7$\%$ of the measured values. The remaining discrepancies between
theory and experiment indicate that inclusion of 1) higher order
corrections and 2) higher-order multipoles contributions, may refine
the differences.

\section*{Acknowledgments}
This work was supported by the research council of the University of
Tehran.

\appendix

\section{Faddeev Equations of $Nd$ scattering in the Doublet and Quartet channels} \label{Appendix A}

\begin{figure}
\includegraphics*[width=13cm]{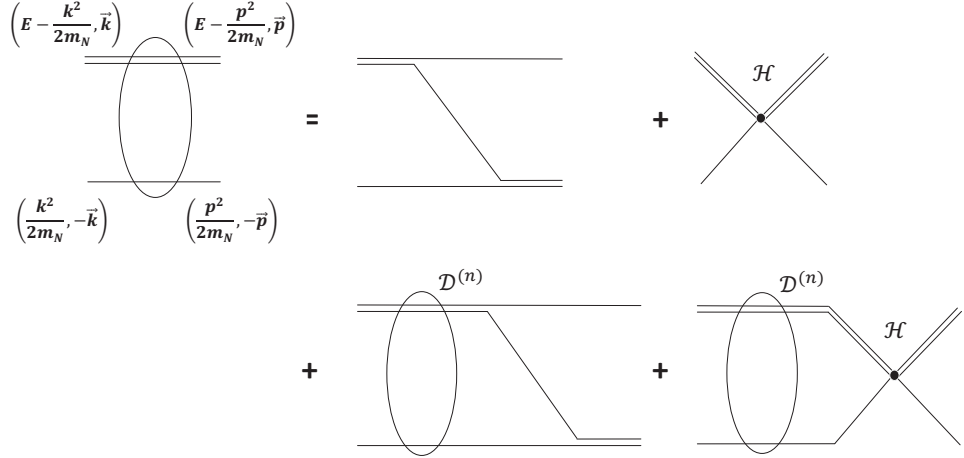}\centering
\caption{\label{Fig:nd scattering}The $Nd$ scattering diagrams up to
$\textrm{N}^n\textrm{LO}$ ($n\leq 2$). All notations are the same as
in Fig.\ref{Fig: nd capture}.}
\end{figure}

The diagrams of the $Nd$ scattering amplitude up to
$\textrm{N}^n\textrm{LO}$ are shown in Fig.\ref{Fig:nd scattering}.
The Faddeev equation of the diagrams in Fig.\ref{Fig:nd scattering}
for the quartet channel in the cluster-configuration space is given
by
\begin{eqnarray}\label{Eq:20}
    \left(
      \begin{array}{cc}
        t^{(n)}_q(E;k,p) & 0 \\
        0 & 0 \\
      \end{array}
    \right)
    = -4\pi \mathcal{K}(E;k,p)\left(
                               \begin{array}{cc}
                                 1 & 0 \\
                                 0 & 0 \\
                               \end{array}
                             \right)
   \qquad\qquad\qquad\qquad\qquad\qquad\quad \nonumber\\+\frac{2}{\pi}\int^\Lambda_0dq\:q^2\:\mathcal{K}(E;q,p)
                                  \mathcal{D}^{
  (n)}(E,q)\left(
      \begin{array}{cc}
        t^{(n)}_q(E;k,p) & 0 \\
        0 & 0 \\
      \end{array}
    \right).\:
\end{eqnarray}
and for the $Nd$ scattering in the doublet ($S$=$\frac{1}{2}$)
channel, we have
\begin{eqnarray}\label{Eq:21}
\left(
  \begin{array}{cc}
    t^{(n)}_{d_{d_tN\rightarrow d_tN}} & t^{(n)}_{d_{d_sN\rightarrow d_tN}} \\
    t^{(n)}_{d_{d_tN\rightarrow d_sN}} & t^{(n)}_{d_{d_sN\rightarrow d_sN}} \\
  \end{array}
\right)(E;k,p) = 2\pi\left[\mathcal{K}(E;k,p)\left(
                                                  \begin{array}{cc}
                                                    1 & -3 \\
                                                    -3 & 1 \\
                                                  \end{array}
                                                \right)
+\mathcal{H}(E,\Lambda)\left(
                 \begin{array}{cc}
                   1 & -1 \\
                   -1 & 1 \\
                 \end{array}
               \right)
                                                            \right]
                                                            \;\nonumber\\-\frac{1}{\pi}\int^\Lambda_0dqq^2\left[\mathcal{K}(E;q,p)\left(
                                                                                                                         \begin{array}{cc}
                                                                                                                           1 & -3 \\
                                                                                                                           -3 & 1 \\
                                                                                                                         \end{array}
                                                                                                                       \right)+\mathcal{H}(E,\Lambda)\left(
                                                                                                                                               \begin{array}{cc}
                                                                                                                                                 1 & -1 \\
                                                                                                                                                 -1 & 1 \\
                                                                                                                                               \end{array}
                                                                                                                                             \right)\right]
                                                                                                                                            \nonumber\\ \times\mathcal{D}^{
  (n)}(E,q)\left(
  \begin{array}{cc}
    t^{(n)}_{d_{d_tN\rightarrow d_tN}} & t^{(n)}_{d_{d_sN\rightarrow d_tN}} \\
    t^{(n)}_{d_{d_tN\rightarrow d_sN}} & t^{(n)}_{d_{d_sN\rightarrow d_sN}} \\
  \end{array}
\right)(E;k,q),\qquad\qquad\;\:
\end{eqnarray}
where $E=\frac{3k^2}{4m_N}-\frac{\gamma^2_t}{m_N}$, $k$ and $p$ are
the total energy of $Nd$ system, the incoming and outgoing
momentums, respectively.
In Eq.(\ref{Eq:21}), $t^{(n)}_{d_{d_xN\rightarrow
d_yN}}$ denotes the $d_xN\rightarrow d_yN$ transition amplitude
($x,y=s$ or $t$) in the doublet channel. The propagator of the
exchanged nucleon, $\mathcal{K}$, is
\begin{eqnarray}\label{Eq:07}
\!\mathcal{K}(E;k,p)=\frac{1}{2}\int^1_{-1}\frac{d
(cos\,\theta)}{\!k^2\!+p^2-ME+kp\:\textrm{cos}\,\theta},
\end{eqnarray}
where $\theta$ indicates the angle between $\vec{k}$ and $\vec{p}$
vectors. Other variables in the above equation are similar to the
text. The results of Eqs.(\ref{Eq:20}) and (\ref{Eq:21}) are
evaluated by considering the operators for projecting the $Nd$
system to $^2S_{\frac{1}{2}}$ and $^4S_{\frac{3}{2}}$ channels. For
the doublet and quartet channels the projection operators
$\mathcal{P}_{d,iA}$ and $\mathcal{P}^j_{q,i}$ are used,
respectively, with iso-spin index $A$ and spin indices $i$ and $j$
\cite{20 of sadeghi-bayegan}.

\section{Triton Wave function} \label{Appendix B}
The normalized triton wave function is obtained by solving the
homogeneous part of Eq.(\ref{Eq:21}) with the application of
$E=-B_t$, where $B_t$ is the binding energy of the triton. So, the
homogeneous part of Eq.(\ref{Eq:21}) for the calculation of the
triton wave function up to $\textrm{N}^n\textrm{LO}$ can be written
as
\begin{eqnarray}\label{Eq:0005}
t^{(n)}_{^3H}(p)=
-\frac{1}{\pi}\int^\Lambda_0dqq^2\Bigg[\mathcal{K}(-B_t;q,p)\left(
                                                                                                                         \begin{array}{cc}
                                                                                                                           1 & -3 \\
                                                                                                                           -3 & 1 \\
                                                                                                                         \end{array}
                                                                                                                       \right)
 \qquad\qquad\qquad\qquad\quad\;\nonumber\\+\mathcal{H}(-B_t,\Lambda)\left(
                                                                                                                                               \begin{array}{cc}
                                                                                                                                                 1 & -1 \\
                                                                                                                                                 -1 & 1 \\
                                                                                                                                               \end{array}
                                                                                                                                             \right)\Bigg]
                                                                                                                                             \mathcal{D}^{
  (n)}(-B_t,q)\,t^{(n)}_{^3H}(q),
\end{eqnarray}
where $t^{(n)}_{^3H}(q)=\bigg(\begin{array}{cc}
    t^{(n)}_{^3H_{d_tN\rightarrow d_tN}}(q) & t^{(n)}_{^3H_{d_sN\rightarrow d_tN}}(q) \\
    t^{(n)}_{^3H_{d_tN\rightarrow d_sN}}(q) & t^{(n)}_{^3H_{d_sN\rightarrow d_sN}}(q) \\
\end{array}\bigg)$. Generally, $t^{(n)}_{^3H_{d_xN\rightarrow
d_yN}}(q)$ denotes the contribution of the ${d_xN\rightarrow d_yN}$
transition ($x,y=s$ or $t$) for making the triton.

One can be able to normalize the solution of Eq.(\ref{Eq:0005}) for
the incoming deuteron channel by \cite{moeini-bayegan}
\begin{eqnarray}\label{Eq:52}
  1=-\int\frac{q^2\,dq}{2\pi^2}\int\frac{q'^2\,dq'}{2\pi^2} \big(t^{(n)}_{^3H}(q)\bigg(\begin{array}{cc}
                                                            1 \\ \,0
                                                          \end{array}\bigg)\big)^\dag \mathcal{D}^{(n)}(-B_t,q)\qquad\qquad\qquad\quad\;\;\;
                                                          \nonumber \\
                                                          \times \frac{\partial}{\partial
                                                          E}\Big[V(E,q,q')\mathcal{D}^{(n)}(E,q')\Big]\bigg|_{E=-B_t}\!\!\!t^{(n)}_{^3H}(q')\bigg(\begin{array}{c}
                                                                                                                                  1 \\
                                                                                                                                  0
                                                                                                                                \end{array}
                                                          \bigg),
\end{eqnarray}
where $V$ is given by
\begin{eqnarray}\label{Eq:53}
  V(E,q,q')=2\pi\left[\mathcal{K}(E;q,q')\bigg(\begin{array}{cc}
                                                                                                                           1 & -3 \\
                                                                                                                           -3 & 1 \\
                                                                                                                         \end{array}
                                                                                                                       \bigg)+\mathcal{H}(E,\Lambda)\bigg(
                                                                                                                                               \begin{array}{cc}
                                                                                                                                                 1 & -1 \\
                                                                                                                                                 -1 & 1 \\
                                                                                                                                               \end{array}
                                                                                                                                             \bigg)\right].
\end{eqnarray}

If we need to find the normalized contribution of the triton wave
function which comes from the incoming singlet dibaryon field, the
replacement $\bigg(\begin{array}{c}
    1 \\
    0
  \end{array}
\bigg)$ by $\bigg(\begin{array}{c}
    0 \\
    1
  \end{array}
\bigg)$ in Eq.(\ref{Eq:52}) must be done.

\end{document}